\documentclass[12pt,preprintnumbers,aps,amssymb,nofootinbib,amsmath]{revtex4}
\usepackage[dvips]{epsfig}
\usepackage{multirow}

\newcommand{\nc}[1]{\newcommand{#1}}

\nc{\its}[1]{\itshape #1 \upshape}
\nc{\mc}[3]{\multicolumn{#1}{#2}{#3}}

\nc{\bc}{\begin{center}}
\nc{\ec}{\end{center}}
\nc{\ig}[1]{\bc \includegraphics{#1} \ec}

\nc{\bo}[1]{\mbox{\boldmath \( #1 \! \! \)  \unboldmath}}
\newcommand{\beqn} {\begin{equation}}
\newcommand{\eqn} {\end{equation}}
\nc{\be}{\begin{eqnarray}}
\nc{\ee}{\end{eqnarray}}
\nc{\bew}{\begin{eqnarray*}}
\nc{\eew}{\end{eqnarray*}}
\nc{\bs}{\begin{subeqnarray}}   
\nc{\es}{\end{subeqnarray}}     
\nc{\nnn}{\nonumber \\}
\nc{\f}[2]{\frac{#1}{#2}}
\nc{\td}[2]{\f{d #1}{d #2}}
\nc{\pd}[2]{\f{\partial #1}{\partial #2}}
\nc{\suli}{\sum\limits}
\nc{\proli}{\prod\limits}
\nc{\ili}{\int\limits}
\nc{\sr}[2]{\stackrel{#1}{#2}}
\nc{\dps}{\displaystyle}
\nc{\ket}[1]{\left| #1 \right>}
\nc{\bra}[1]{\left< #1 \right|}
\nc{\bracket}[2]{\left< #1 \right| \left. \! #2 \right>}
\nc{\norm}[1]{\left\| #1 \right\|}
\nc{\lndm}[1]{\pd{^{#1} \ln{\det{M}}}{\mu^{#1}}}
\nc{\pdmm}[1]{M^{-1} \pd{^{#1} M}{\mu^{#1}}}
\nc{\pdm}{M^{-1}\pd{M}{\mu}}
\nc{\trac}[1]{\mbox{Tr}\left(#1\right)}
\nc{\hm}{\hat{m}}
\nc{\hr}{\hat{r}}
\newcommand{\tr}{{\rm Tr}}

\def\lsim{\raise0.3ex\hbox{$<$\kern-0.75em\raise-1.1ex\hbox{$\sim$}}}
\def\gsim{\raise0.3ex\hbox{$>$\kern-0.75em\raise-1.1ex\hbox{$\sim$}}}

\begin{document}

\title{Correlation functions of the energy-momentum tensor \\
in SU(2) gauge theory at finite temperature}

\author{K. H\"ubner$^a$, F. Karsch$^{a,b}$ and C. Pica$^a$}

\affiliation{
$^{\rm a}$Physics Department, Brookhaven National Laboratory, 
Upton, NY 11973, USA \\
$^{\rm b}$Fakult\"at f\"ur Physik, Universit\"at Bielefeld, D-33615 Bielefeld,
Germany
}

\date{\today}
\preprint{BNL-NT-08/9}
\preprint{BI-TP 2008/08}

\begin{abstract}
We calculate correlation functions of the energy-momentum tensor
in the vicinity of the deconfinement phase transition of (3+1)-dimensional
SU(2) gauge theory and discuss their critical behavior in the vicinity of
the second order deconfinement transition. We show that correlation
functions of the trace of the energy momentum tensor diverge uniformly
at the critical point in proportion to the specific heat singularity.
Correlation functions of the pressure, on the other hand, stay finite
at the critical point. We discuss the consequences of these findings 
for the analysis of transport coefficients, in particular the bulk
viscosity, in the vicinity of a second order phase transition point.
\end{abstract}

\pacs{11.15.Ha, 11.10.Wx, 12.38Gc, 12.38.Mh}

\maketitle

\section{Introduction}
\label{intro}
Experimental evidence for rapid thermalization of the dense
matter created in heavy ion collisions at RHIC \cite{RHIC} provided strong
motivation for a new look at the non-perturbative structure of the quark gluon
plasma (QGP).
The rapid thermalization of the QGP has been related to the 
smallness of the shear viscosity in the plasma phase \cite{Teaney}, which may
come close to the theoretical lower bound established for the ratio of 
shear viscosity to entropy density \cite{Son}. This gave rise to the 
interpretation
of the quark gluon plasma above but close to the transition temperature as a
strongly interacting medium that has properties of an almost perfect liquid.

These experimental findings also renewed the interest in determining 
transport properties
of gauge theories through the calculation of correlation functions of the 
energy-momentum tensor on the lattice. This is by far not a straightforward
calculation and requires a careful analysis of the low frequency structure
of the spectral representation of these correlation functions \cite{Wyld}.
Nonetheless progress has been made through high statistics calculations
in SU(3) gauge theories \cite{Nakamura,Meyer_shear,Meyer_bulk}.

Recently it has been argued that close to the transition from low temperature
hadronic matter to the plasma phase of QCD bulk viscosity might play a
much more important role than shear viscosity \cite{Kharzeev}; 
bulk viscosity has been related to  a temperature derivative of the
trace of the energy-momentum tensor. The latter does diverge at a second order
critical point \cite{Tuchin} and thus would force bulk viscosity to diverge too.
This argument is based on a sum rule 
derived for correlation functions of the energy-momentum tensor in 
Minkowski space \cite{Kharzeev}. 
However, as noted recently \cite{Meyerc}, in Euclidean 
space the correlation function of the trace of the energy-momentum tensor 
picks up an additional constant contribution corresponding to an exact 
$\delta$-function at zero frequency in the spectral representation of this 
correlator. This contribution, in fact, cancels the leading singular
behavior of the correlation function and thus
makes the relation between transport coefficients and properties of 
bulk thermodynamic observables more subtle. 

In the following we will provide evidence for the presence of a contribution 
to the correlation function of the trace of energy-momentum
tensor that is constant in Euclidean time but temperature dependent. We show
that the temperature dependence of this contribution scales like the specific
heat and thus dominates the singular behavior of this correlation function
in the vicinity of $T_c$.
These findings also explain the 
inconsistency observed in the parametric dependence of bulk viscosity
derived in high temperature perturbation theory and through the sum rule
analysis \cite{Moore}. 
 
The singular behavior of bulk viscosity in the vicinity of a critical point has
long been known in statistical physics \cite{Fixman,Kawasaki}. In particular,
the divergence of bulk viscosity at the critical point
of the liquid gas transition has been analyzed extensively \cite{Hohenberg}.
In this case the divergence of bulk viscosity ($\zeta$) has been related 
to the critical behavior
of the 3-dimensional Ising model \cite{Kawasaki,Kadanoff,Onuki}, which is in
the universality class for the critical point of the liquid-gas phase 
transition. It has been argued that 
the divergence of $\zeta$ is strong and, in fact, almost quadratic in the 
inverse reduced temperature, $t=|T-T_c|/T_c$. The singular 
behavior, $\zeta \sim t^{-z\nu+\alpha}$, with $\alpha,\;\nu$ denoting static 
critical exponents of
the 3-d Ising model and $z$ being a dynamical exponent characterizing the
equilibration of density fluctuations, goes along with a strong divergence 
of the relaxation time for density fluctuations, $\tau_R$. Their ratio,
$\zeta/\tau_R \sim t^\alpha$, however, is proportional to the inverse of
the specific heat and thus vanishes slowly at the critical point. 

While numerous numerical studies have confirmed that the critical behavior of
bulk thermodynamics in $(d+1)$-dimensional gauge 
theories is indeed controlled by $d$-dimensional universality classes 
corresponding to the global symmetry of the relevant order parameter, little is
known about the relation between dynamic properties and the related
critical exponents \cite{Schuelke}. 
The SU(2) gauge theory with its second order deconfinement phase transition
seems to be an ideal model to explore critical behavior of dynamical
properties, e.g. transport coefficients.
The SU(2) gauge theory at finite temperature incorporates all the basic 
features one expects to be relevant for deconfinement in QCD. Moreover, 
it has a second order phase transition which
belongs to the universality class of the 3-dimensional Ising model. It
thus will allow to analyze in how far dynamic universal properties known
for the Ising universality class are relevant for transport properties
in a quantum field theory. It also may give insight into transport properties 
in the vicinity of the chiral critical point in QCD that may exist at 
non-zero baryon number density and also would belong to the Ising universality 
class \cite{Stephanov}. 
In the vicinity of the SU(2) deconfinement phase transition 
a lot of information exists about the critical behavior
of bulk thermodynamics that can be related to the critical behavior
of correlation functions of the energy-momentum tensor. This will help to
gain experience with lattice calculations of the bulk viscosity in pure
gauge theories. 

In this paper we will establish relations between the critical behavior
of bulk thermodynamic observables in the vicinity of
the SU(2) deconfinement transition and properties of correlation functions
of the energy-momentum tensor.
In particular, we will analyze the 
Euclidean correlation functions of the trace of the energy-momentum tensor 
and compare it to critical behavior of the specific heat. 
We present evidence that the strength of the divergence of this correlation 
function at $T_c$ is independent of Euclidean time and is controlled by the 
universal structure of the divergence of the specific heat at a critical 
point belonging to the 3-dimensional Ising universality class.
In the next section we review the calculation of bulk thermodynamic
observables in SU(2) gauge theory \cite{Redlich} and present some new
results on the diagonal components of the energy-momentum tensor,
{\it i.e.} energy density and pressure, in the vicinity of the deconfinement
transition temperature. In Section III we introduce local operators for 
energy density and pressure and discuss basic properties of their finite 
temperature 
Euclidean time correlation functions at vanishing momentum. Section IV
is devoted to a discussion of critical behavior extracted from correlation
functions of the trace anomaly and presents a calculation of the
pressure-pressure correlator. We analyze consequences for the determination
of bulk viscosity from these correlation functions in Section V. Finally, we 
conclude in Section VI.

\section{Thermodynamics of SU(2) gauge theories on the lattice}

As preparation for the analysis of correlation functions of 
diagonal elements of the energy-momentum tensor
we want to discuss here the calculation of the relevant bulk thermodynamic 
observables, energy density ($\epsilon$) and pressure ($P$), 
on the lattice. In order to analyze Euclidean space-time
correlation functions of $\epsilon$, $P$ or the trace of the 
energy-momentum tensor, $\Theta^{\mu\mu}=\epsilon - 3P$, one obviously
needs to use an approach to bulk thermodynamics that allows to define
local operators for energy density and pressure. We will call this approach,
which in fact was the 
basis for the original formulation of bulk thermodynamics on the lattice
\cite{Engels}, the differential formalism. It is in contrast to the 
so-called integral method which nowadays is most commonly used in lattice
calculations to extract
energy density and pressure from the trace anomaly \cite{Miller}.
The differential formalism has not been used recently in studies  
of bulk thermodynamic observables as it is more 
involved and requires additional non-perturbative calculations of 
derivatives of the bare gauge couplings with respect to the anisotropy
parameter, $\xi=a_\sigma/a_\tau$, that controls the  
spatial ($a\equiv a_\sigma$) and temporal ($a_\tau$) lattice
spacings \cite{Karsch}. Having control over the anisotropy is needed 
to derive thermodynamic quantities as function of temperature, 
$T=1/N_\tau a_\tau$, and volume, $V=(N_\sigma a_\sigma)^3$ through 
appropriate partial derivatives with respect to $a_\tau$ and $a_\sigma$ 
\cite{Engels}, although numerical calculations will generally be performed on 
isotropic lattice ($\xi =1$).
In particular, for the study of the singular behavior of the energy density
in the vicinity of the SU(2) deconfinement transition this approach
has been used successfully \cite{Redlich}. In this context also the necessary  
derivatives of the gauge coupling with respect to $\xi$ had been
analyzed in detail. We will follow this analysis here closely.

\subsection{The differential formalism}

We start from the partition function for a $SU(N_c)$ gauge theory at
finite temperature, which we write in standard lattice notation \cite{review},
\begin{equation}
Z(T,V)= \int \prod_{x,\mu} {\rm d}U_{x,\mu}\;
 {\rm e}^{-S(T,V)} \; ,
\label{partition}
\end{equation}
with
\begin{equation}
S(T,V)  = \frac{2N_c}{g^2} \sum_{x,\mu,\nu} \left(1-P_{\mu\nu}(x)\right) \; . 
\label{action}
\end{equation} 
Here $x=(\vec{x},x_4)$ labels the discrete set of space-time points on a 
four-dimensional hypercubic lattice of size $N_\sigma^3N_\tau$.

In the standard Wilson discretization scheme the local gauge action in the
$\mu\nu$-hyperplane is expressed in terms of plaquette variables
$P_{\mu\nu}(x)\equiv \frac{1}{N_c} {\rm Re} \tr 
U_{x,\mu}U_{x+\hat{e}_\mu,\nu}U^\dagger_{x+\hat{e}_\nu,\mu}U^\dagger_{x,\nu}$, 
with link variables $U_{x,\mu} \in SU(N_c)$.
We write the action in terms of contributions arising only from 
entirely space-like plaquettes ($P_\sigma$) and space-time like
plaquettes ($P_\tau$),
\begin{eqnarray}
P_\sigma = \frac{1}{N_\sigma^3 N_\tau} \sum_x P_\sigma (x) \;\; &,&\;\; 
P_\sigma (x) = \frac{1}{3} \sum_{i>j=1,2,3} P_{ij}(x)\;\; , \nonumber \\ 
P_\tau =  \frac{1}{N_\sigma^3 N_\tau} \sum_x P_\tau (x) \;\; &,&\;\;
P_\tau(x) = \frac{1}{3} \sum_{j=1,2,3}P_{4,j} (x)  \; .
\label{plaquettes}
\end{eqnarray}
We also introduce $P_0=(P_\sigma+P_\tau)/2$ as the average of 
space-like and time-like
plaquettes evaluated on a symmetric ($T\simeq 0$) lattice, {\it i.e.}
for $N_\tau=N_\sigma$. 

From the logarithm of the partition function one obtains pressure and
energy density as derivatives with respect to $a_\sigma$ and $a_\tau$,
respectively \cite{Engels},
\begin{eqnarray}
\frac{P}{T^4} &=&
 N_c N_\tau^4\left( \left( 2g^{-2}-
\left( c_\sigma - c_\tau \right) \right)
  (P_{\sigma}-P_{\tau}) 
    -3 \left( c_{\sigma}+c_{\tau}  \right)
  \left( 2P_0-(P_{\sigma}+P_{\tau}) \right)
       \right)    \; ,  \label{thermoP} \\
\frac{\epsilon}{T^4} &=&
 3 N_c N_\tau^4\left( \left( 2g^{-2}-
\left( c_\sigma - c_\tau \right) \right)
  (P_{\sigma}-P_{\tau}) 
    + \left( c_{\sigma}+c_{\tau}  \right)
  \left( 2P_0-(P_{\sigma}+P_{\tau}) \right)
       \right) \; .
\label{thermo}
\end{eqnarray} 
Here $c_{\sigma}$ and $c_{\tau}$ are given in terms of derivatives of
the gauge coupling, $g^{-2}(a,\xi)$, with respect to the anisotropy at $\xi =1$
\cite{Engels,Karsch}. Their sum is constraint by the $\beta$-function,
\begin{equation}
B(g^{-2}) =  a\frac{{\rm d} g^{-2}}{{\rm d}a} \biggr|_{\xi=1}= -2 
\left( c_{\sigma}+c_{\tau} \right) \; .
\label{beta}
\end{equation}

With these relations for $\epsilon$ and $P$ it is straightforward to obtain 
also the trace of the energy-momentum tensor and the entropy density,
\begin{eqnarray}
\frac{\Theta^{\mu\mu}}{T^4} &\equiv& 
\frac{\epsilon-3p}{T^4} =
6 B(g^{-2}) N_c N_\tau^4 \; \left[ P_{\sigma}+P_{\tau} -2P_0\right] \;, 
\label{e3pP}\\
\frac{s}{T^3} &\equiv& \frac{\epsilon+p}{T^4} = 
 4 N_c N_\tau^4\left( \left[ 2g^{-2}-
\left( c_\sigma - c_\tau \right) \right]
  (P_{\sigma}-P_{\tau}) \right)
\; .
\label{e3p}
\end{eqnarray} 
We note that these particular combinations of energy density and 
pressure are proportional to simple sums and
differences of space-like and time-like plaquette expectation values. They
only require common multiplicative renormalizations for $P_\sigma$ and
$P_\tau$. Extracting $\epsilon/T^4$ or $P/T^4$ directly, however, involves
different renormalization factors for space-like and time-like plaquette
expectation values. For this reason it is straightforward to analyze 
correlation functions of local densities of the entropy or trace anomaly.
We note that in the vicinity of the SU(2) deconfinement transition the 
dominant singular behavior of $\Theta^{\mu\mu}/T^4$ as well as $s/T^3$ is 
controlled by the energy density. We will show later that in the 
vicinity of $T_c$ the pressure is an order of magnitude smaller than
the energy density. 
When one combines $\Theta^{\mu\mu}/T^4$ and $s/T^3$ to extract the pressure 
it thus requires good knowledge of the renormalization factors
$c_\sigma$ and $c_\tau$ in order to eliminate the dominant energy 
contributions and arrive at a proper definition for 
a local operator for the pressure. This is a prerequisite for the analysis
of pressure-pressure correlation functions and, as
will become clear later, this becomes of importance for studies of 
transport coefficients, e.g. the bulk viscosity, in the vicinity of
$T_c$.

\subsection{Critical energy density and pressure}

As an application of the differential formalism reviewed above we have
performed a finite size scaling analysis of the critical energy density 
and the pressure at the deconfinement transition temperature of the SU(2) 
gauge theory. This extends earlier studies of the critical energy density
performed in Ref.~\cite{Redlich} to larger lattices and yields, for the
first time, a determination of the pressure at the critical point.

The deconfinement transition of the SU(2) gauge theory in (3+1)-dimensions
is in the same universality class as the 3-dimensional Ising model. Its 
critical behavior is well understood and has been confirmed in various
finite size scaling studies. On a lattice of size $N_\sigma^3 N_\tau$
energy density and pressure at the critical point are expected to scale as
\begin{eqnarray}
\left( \frac{P(T_c)}{T_c^4} \right)_{N_\tau, N_\sigma} &=& 
\left( \frac{P(T_c)}{T_c^4} \right)_{N_\tau, \infty} + 
a_P \left(\frac{N_\tau}{N_\sigma}\right)^{3} \; ,
\label{Pc} \\
\left( \frac{\epsilon(T_c)}{T_c^4} \right)_{N_\tau, N_\sigma} &=& 
\left( \frac{\epsilon(T_c)}{T_c^4} \right)_{N_\tau, \infty} + 
a_\epsilon \left(\frac{N_\tau}{N_\sigma}\right)^{(1-\alpha)/\nu} \; .
\label{ec}
\end{eqnarray} 
Here $\alpha = 0.110(1)$ and $\nu = 0.6301(4)$ denote the critical exponents of the
3-dimensional Ising universality class, controlling the divergence of the 
specific heat and the correlation length, 
respectively\footnote{Note that the leading finite volume correction that
arises from the singular contribution to the 
pressure is inverse proportional to the volume as it is the case also
for the leading corrections to regular terms. This is due to
a hyper-scaling relation, $2-\alpha=d\nu$ with $d=3$.}. 
\begin{figure}[t]
\begin{center}
\epsfig{file=avr_ep.eps, width=7.8cm}\hspace*{0.2cm}
\epsfig{file=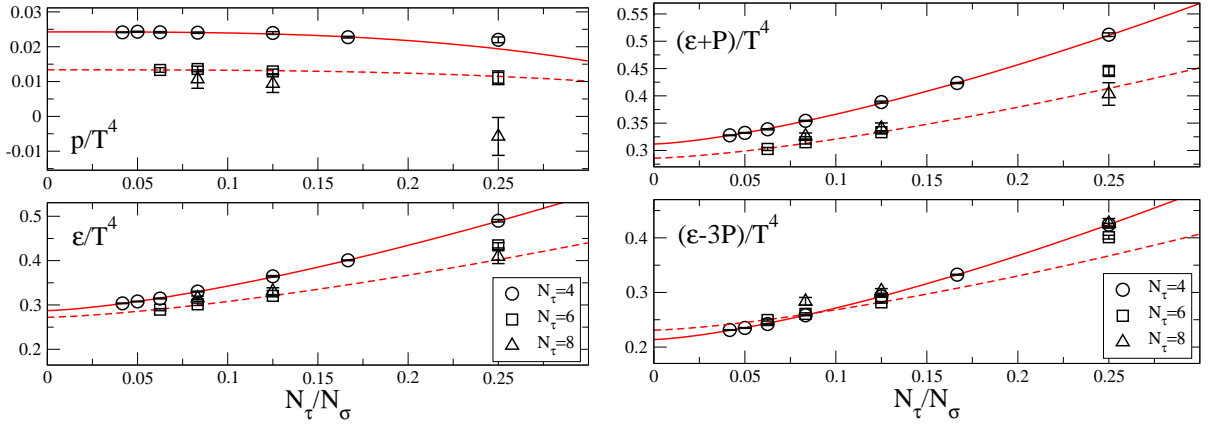, width=7.8cm}
\end{center}
\vspace*{-0.4cm}
\caption{Finite size scaling analysis of bulk thermodynamic observables
on lattices of size $N_\tau=4,\; 6$ and $8$. The Curves show fits that
have been performed for $N_\tau/N_\sigma \le 1/6$.
}
\label{fig:bulkscaling}
\end{figure}

We have analyzed the scaling behavior of energy density and pressure
on lattices with temporal extent $N_\tau = 4,\; 6$ and $8$ for various spatial 
lattice sizes; covering the range $N_\sigma \in [16,96]$ for $N_\tau=4$,
and $N_\sigma \in [24,96]$ for $N_\tau=6$ and $N_\sigma \in [32,96] $ for 
$N_\tau=8$. For these values of $N_\tau$ the infinite volume critical 
couplings are known quite precisely \cite{Engels:1998nv,Engels:1992fs,Fingberg}. 
We have listed them in Table~\ref{tab:critical}. 
At these values of the gauge coupling also the 
anisotropy coefficients $c_\sigma$ and $c_\tau$ have been determined
previously \cite{Redlich}. 

We have calculated pressure, energy density, entropy density and
trace anomaly on lattices of temporal extent $N_\tau =4,\; 6$ and $8$ 
for various spatial lattice sizes at the critical couplings given in
Table~\ref{tab:critical}. 
Data for the vacuum subtraction needed for all observables but the entropy
density have been obtained through simulations on a $32^4$ lattice. 
The number of configurations analyzed ranges from $1\cdot 10^{6}$ on our 
smaller lattices to $4\cdot 10^{5}$ on the large lattices. 
For our calculations on $N_\tau=4$ and $6$ lattices we have used
a standard heat-bath/over-relaxation algorithm. 
For calculations on the larger $N_\tau=8$ lattices it became possible
to use also a two-level algorithm \cite{luscher} which turned out to be 
more efficient for the analysis of correlation functions. 
Autocorrelation times
in the transition region range from ${\cal O}(1)$ on the small lattices
to about $40$ on the largest lattices.

In Fig.~\ref{fig:bulkscaling} we show results for the
volume dependence of various bulk thermodynamic observables. It 
clearly can be seen that the scaling behaviour of the pressure is different
from that of the other three observables. Moreover, it is apparent that the pressure
is more than an order of magnitude smaller than any of the observables that
contain contributions of the energy density. This shows that the finite
size scaling behavior of $(\epsilon + P)/T^4$ and $(\epsilon - 3P)/T^4$ is
dominated by the non-analytic behavior of the energy density and thus is
expected to be controlled by the same ansatz as for the energy density 
given in Eq.~\ref{ec}. We also note that cut-off effects seem to be 
small already for lattices with temporal extent $N_\tau\gsim 6$ and that 
the volume dependence starts being controlled by the scaling variable 
$V^{1/3}T=N_\sigma/N_\tau$. This may be expected as bulk thermodynamics in 
the vicinity of $T_c$ is dominated by low frequency modes for which the 
underlying lattice cut-off becomes unimportant.

To check consistency with the scaling behavior of the Ising universality class
we have fitted the data for the energy on the $N_\tau=4$ lattice for all
$N_\sigma/N_\tau > 4$ to the ansatz
given in Eq.~\ref{ec} with a free scaling exponent $\Delta=(1-\alpha)/\nu$.
This fit yields $\Delta=1.41(6)$ which is in good agreement with the value
in the 3-dimensional Ising universality class, $\Delta = 1.412(1)$ 
\cite{Pelissetto:2000ek}. Fits to the finite volume corrections for the 
pressure yield an exponent $d=3.0\pm 1.4$ in good agreement with the 
expected hyper-scaling relation $d=3$.

To extract the asymptotic, infinite volume 
values for the critical energy density and the pressure we then used the
ans\"atze given in Eqs.~\ref{Pc} and \ref{ec} with fixed scaling exponents
$1.412$ and $3$, respectively.
Results for $\epsilon(T_c)/T_c^4$ and $P(T_c)/T_c^4$ for 
$N_\sigma/N_\tau =12$ as well as infinite volume extrapolated
values are summarized in 
Table~\ref{tab:critical}; results for the critical energy density are 
consistent with those of Ref.~\cite{Engels} but have a statistical error 
that is an order of magnitude smaller. Comparison of the $N_\tau =8$
results with those for $N_\tau=6$ suggest that the remaining cut-off 
dependence is small. 

\begin{table}[t]
\begin{center}
\vspace{0.3cm}
\begin{tabular}{|c|c||l|l||l|l|}
\hline
~$N_\tau$~&~$\beta_c$~&
~$\epsilon(T_c)/T_c^4$~&~$P(T_c)/T_c^4$& 
~$\epsilon(T_c)/T_c^4$~&~$P(T_c)/T_c^4$~ \\
\hline
~&~&\multicolumn{2}{c||}{$N_\sigma=12 N_\tau$}&
\multicolumn{2}{|c|}{$N_\sigma\rightarrow\infty$}\\
\hline
~4~ & ~2.29895(10)~ & ~0.3303(9)~&~0.0240(2)~&~0.2872(5)~ & ~0.0242(3)~ \\
~6~ & ~2.4265(30)~ & ~0.3015(27)~&~0.0136(8)~&~0.2722(31)~ & ~0.0135(8)~ \\
~8~ & ~2.5115(40)~ & ~0.3158(42)~&~0.0107(26)&~~~~~--~ & ~0.0107(15)~ \\
\hline
\end{tabular}
\end{center}
\caption{Critical energy density and pressure calculated on lattices with
temporal $N_\tau$ on lattices with aspect ratio $N_\sigma/N_\tau =12$ and 
infinite volume extrapolated results.
The second column gives the values of the gauge couplings
at which these calculations have been performed. The error estimates are 
based on fits with the ans\"atze given in Eqs.~\ref{ec} and \ref{Pc} and does
not include errors that arise from uncertainties in the couplings 
$c_\sigma$, $c_\tau$ as well as the critical coupling $\beta_c$.
}
\label{tab:critical}
\end{table}

\section{Correlation functions of the energy-momentum tensor} 

Within the differential framework for bulk thermodynamics on the
lattice, reviewed in the previous section, it is straightforward 
to define local operators for the energy density and pressure, as 
well as the entropy density and the trace of the energy momentum
tensor. Using $P_\tau(x)$ and $P_\sigma(x)$ as introduced in 
Eq.~\ref{plaquettes} one obtains from Eqs.~\ref{thermoP},~\ref{thermo}
and Eqs.~\ref{e3pP},~\ref{e3p}
local operators for pressure, energy and entropy density as well as
the trace anomaly, e.g.
\begin{equation}
\Theta^{\mu\mu}(\vec{x},x_4) = 
6 N_c B(g^{-2}) \; \left[ P_{\sigma}(\vec{x},x_4)+P_{\tau}(\vec{x},x_4) -
2P_0\right] \;
\label{Thetax}
\end{equation}
for the trace anomaly and similar for other observables. The
local plaquette variables, $P_{\sigma}(\vec{x},x_4),\; P_{\tau}(\vec{x},x_4)$
have been defined in Eq.~\ref{plaquettes}.
We will denote this particular choice of local expressions for plaquette
variables in the following as 
discretization scheme 1. However, when considering correlation
functions in Euclidean time, one may want to account for the effect
that time-like plaquettes connect two different time-slices in a lattice
and should be connected to a Euclidean time that is displaced from that
of space-like plaquettes.
This ambiguity arises because gauge fields are introduced
on links of the lattice and the field strength tensor thus is spread
at least over a domain of the size of an elementary plaquette. This effects 
introduce additional ${\cal O}(a^2)$ discretization errors, {\it i.e.}
errors that are of the same order as the discretization errors that exist 
anyhow in the definition of the Euclidean action or the energy-momentum 
tensor used in our calculations. They, of course, will disappear in the 
continuum limit. Nonetheless, comparing different discretization schemes
will allow to estimate the magnitude of systematic ${\cal O}(a^2)$ effects.
We therefore have introduced two other discretization schemes which 
symmetrize the contribution of space-like plaquettes relative to a 
time-like plaquette or vice verse,
\begin{center}
scheme 2: $P_{ij}(x) \rightarrow 
\frac{1}{2}\left( P_{ij}(x)+P_{ij}(x+\hat{e}_4) \right)$  , \\ 
scheme 3: $P_{0j}(x) \rightarrow 
\frac{1}{2}\left( P_{4j}(x)+P_{0j}(x-\hat{e}_4) \right)$  .
\end{center}
Note that there is no need for a local definition for $P_0$ which enters the
definition of $\Theta^{\mu\mu}(x)$.
$P_0$ is evaluated at zero temperature and is a constant, identical
at all space-time points. We furthermore introduce zero-momentum projected
operators and their fluctuations,
\begin{eqnarray}
Y(\tau) &=& \frac{1}{N_\sigma^3} \sum_{\vec{x}} Y(\vec{x},\tau) \; ,
\nonumber\\ 
\bar{Y}(\tau) &=& Y(\tau) -\langle \frac{1}{N_\tau}
\sum_{\tau =1}^{N_\tau}Y(\tau)\rangle \; .
\label{operator}
\end{eqnarray}
Correlation functions are then obtained as thermal averages over products of 
fluctuation operators, {\it i.e.} we consider connected correlation functions,
\begin{equation}
\frac{G_{XY}(\tau,T)}{T^5} = N_\tau^5 \langle \bar{X}(\tau) \bar{Y}(0) 
\rangle \; .
\label{correlator}
\end{equation}
Due to the periodic boundary conditions in Euclidean time
points in the hyperplane at $\tau=0$ are identical to those at
$\tau=N_\tau$.
In the following we also will discuss properties of the midpoint-subtracted
correlation functions,
\begin{equation}
\frac{\Delta G_{XY}(\tau,T)}{T^5} = N_\tau^5 \left( 
\langle {X}(\tau) {Y}(0) \rangle -
\langle {X}(1/2T) {Y}(0) \rangle \right) \; .
\label{dif_correlator}
\end{equation}

\begin{figure}[t]
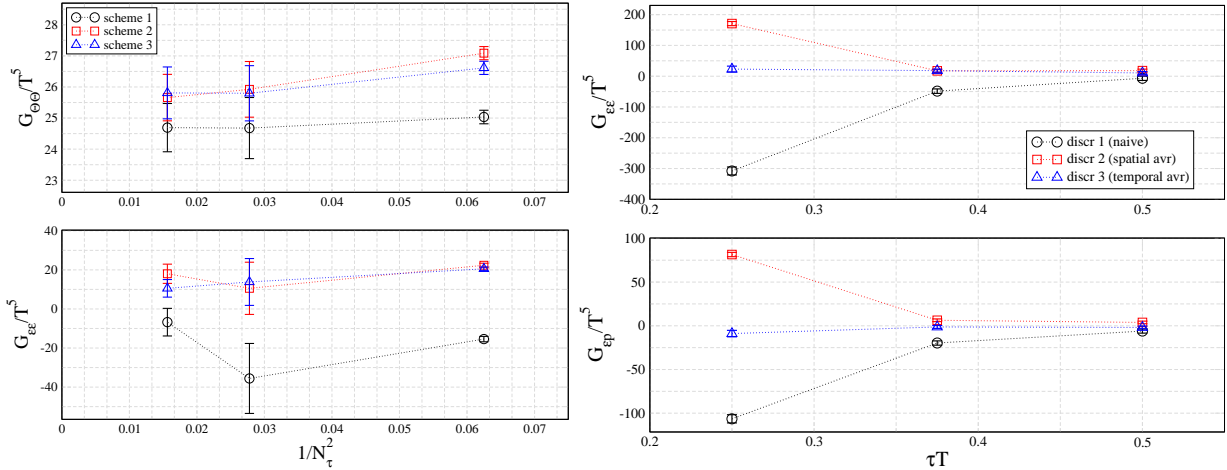

\begin{center}
\epsfig{file=discr_ntdep_comb.eps, width=7.5cm}
\epsfig{file=discr_ecorr_nt8.eps, width=8.6cm}
\end{center}
\caption{Cut-off dependence in different discretization schemes of the
energy momentum tensor. The left hand part of the figure shows results
for the trace-trace and energy-energy correlation functions for different
values of the lattice cut-off $aT=1/N_\tau$ evaluated on lattices with 
spatial extent $N_\sigma = 12 N_\tau$.
The right hand part of the figure shows results for the energy-energy
and energy-pressure correlation functions versus Euclidean time, $\tau T$.
}
\label{fig:cutoff}
\end{figure}

We used the three discretization schemes introduced above for diagonal
components of the energy-momentum tensor
to analyze the cut-off dependence of correlation functions. In
Fig.~\ref{fig:cutoff}(left) we show results for correlation functions of
the trace of the energy momentum tensor, $G_{\Theta\Theta}$
and the energy, $G_{\epsilon\epsilon}$. 
These correlators have been evaluated on lattices with the largest
temporal extent, $N_\tau = 8$, used in this study, and 
at the largest Euclidean
time separation possible at finite temperature, i.e. the midpoint of 
the temporal direction $\tau =1/2T$. It is apparent from these figures that 
cut-off effects in $G_{\Theta\Theta}$ are only of the order of 10\%;
within the current errors different discretization schemes yield consistent 
results on the $N_\tau=8$ lattices. For $G_{\epsilon\epsilon}$, however,
discretization errors are large even on the largest lattices.

\begin{figure}[t]
\begin{center}
\epsfig{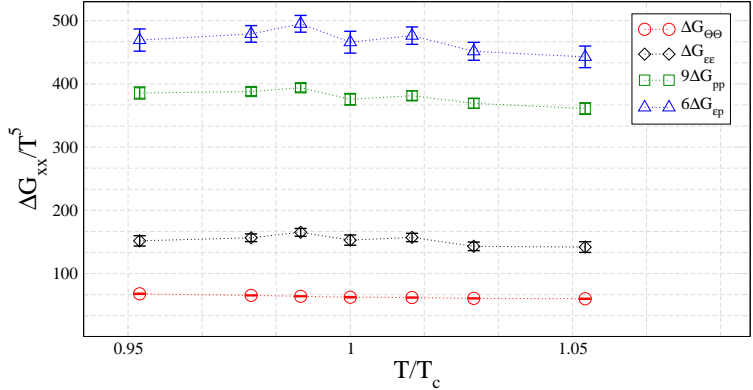}
\end{center}
\caption{Midpoint-subtracted correlation functions on lattices with
temporal extent $N_\tau=8$ versus temperature. Shown are various 
contributions to the correlation function of the trace anomaly
(Eq.~\ref{D_decomposed}). 
}
\label{fig:cutoff2}
\end{figure}

In the right hand part of Fig.~\ref{fig:cutoff} we show the energy-energy and 
energy-pressure correlation functions as function of Euclidean time, $\tau T$,
evaluated at the critical temperature. As has been pointed out in 
Ref.~\cite{Meyerc} these correlation functions will, in the continuum limit,
be independent of Euclidean time as the energy operator is time independent,
\begin{equation}
G_{\epsilon X}(\tau, T) = T^2 \frac{\partial}{\partial T} \langle X(0)\rangle 
\; .
\label{exact}
\end{equation}
At finite lattice spacing, however, this is not yet the case. It is apparent
from Fig.~\ref{fig:cutoff}(right) that the influence of a finite lattice
cut-off is more severe at shorter distances, and 
correlation functions $G_{\epsilon X}$ thus show a significant dependence
on Euclidean time.
As a consequence, the simple relation between midpoint-subtracted
correlation functions of $\Theta^{\mu\mu}$ and pressure-pressure correlators,
\begin{eqnarray}
\Delta G_{\Theta\Theta} (\tau,T) &=& \Delta G_{\epsilon\epsilon} (\tau,T) + 
9 \Delta G_{PP} (\tau,T)
-6 \Delta G_{\epsilon P} (\tau,T) \label{D_decomposed} \\
&\equiv& 9 \Delta G_{PP} (\tau,T) + {\cal O}(a^2) \; ,
\label{continuum_decomposed}
\end{eqnarray} 
which holds in the continuum limit, is violated by ${\cal O}(a^2)$ corrections
at finite lattice spacing.
The left hand side of Eq.~\ref{D_decomposed} as well as the various components
of the right hand side are shown in Fig.~\ref{fig:cutoff2}
for $\tau T=1/4$ and $N_\tau =8$. In
the continuum limit the differences $\Delta G_{\epsilon\epsilon} (\tau,T)$
and $\Delta G_{\epsilon P} (\tau,T)$ should vanish and 
$\Delta G_{\Theta\Theta} (\tau,T)$ should be equal to 
$9 \Delta G_{PP} (\tau,T)$.
This is clearly not the case for finite values of the lattice cut-off.

\section{Correlation functions and critical behavior}

Fortunately, the cut-off dependence discussed in the previous section 
is not crucial for the analysis of critical behavior as it 
arises from short-distance effects. Close to $T_c$, however, the 
correlation length is large and thermal effects are not very sensitive to
the underlying cut-off. The cut-off dependence thus 
is part of the smooth regular background that contributes to $G_{XY}(\tau,T)$.  
This is apparent from the weak temperature dependence of the differences 
$\Delta G_{XY}(\tau,T)$ shown in Fig.~\ref{fig:cutoff2}. 
Moreover, we find that correlators calculated in the three 
discretization schemes show a similar temperature dependence and differ, at 
fixed $\tau T$ only by an almost temperature independent shift.

\subsection{Correlation function of \boldmath$\Theta^{\mu\mu}$}

We will discuss here the temperature dependence of correlation functions
of $\Theta^{\mu\mu}(\vec{x},\tau)$ at vanishing momentum, {\it i.e.} we
calculate $G_{\Theta\Theta} (\tau,T)/T^5$ as introduced in Eq.~\ref{correlator}.
This correlation function is the sum of three contributions
corresponding to energy-energy, pressure-pressure and energy-pressure 
correlations, respectively,
\begin{equation}
G_{\Theta\Theta} (\tau,T) = G_{\epsilon\epsilon} (\tau,T) + 9 G_{PP} (\tau,T)
-6 G_{\epsilon P} (\tau,T)\; .
\label{decomposed}
\end{equation}
As discussed above in the continuum limit the
entire $\tau$-dependence of $G_{\Theta\Theta}$ is 
expected to arise from the pressure-pressure correlations, 
while close to $T_c$ the dominant temperature dependence will arise from the
energy-energy correlator, which through the fluctuation-dissipation theorem,
is proportional to the specific heat,
\begin{equation}
\frac{G_{\epsilon\epsilon} (\tau,T)}{T^5} \sim \frac{{\rm d}\epsilon}{{\rm d}T}
\sim \frac{c_V}{T^3} \; .
\label{efluctuation}
\end{equation}
Close to the deconfinement transition we therefore expect that 
$G_{\Theta\Theta} (\tau,T)$ will show critical behavior that 
coincides with that of the specific heat in a 3-dimensional Ising model,
\begin{equation}
\frac{G_{\Theta\Theta} (\tau,T)}{T^5}\; \sim \;   \frac{c_V}{T^3} \;
\sim \; A_{\pm}
\left| \frac{T-T_c}{T_c} \right|^{-\alpha}\left( 1+ B_\pm 
\left| \frac{T-T_c}{T_c} \right|^{\omega} + ...\right) 
\quad
{\rm for} \quad T\rightarrow T_c^\pm \; .
\label{Gcrit}
\end{equation}
Here $\alpha$ is the specific heat critical exponent and $\omega$ is the
correction to scaling exponent which characterizes the leading
non-analytic scaling correction. Like $\alpha$ also the exponent $\omega$
is universal. For the $SU(2)$ gauge
theory in (3+1)-dimensions, which we are analyzing here, the relevant
universality class is that of the 3-dimensional Ising model. In this case
$\alpha = 0.110(1)$ and $\omega=0.53(3)$ \cite{Pelissetto:2000ek} which 
also has been verified in lattice calculations \cite{Redlich}. 

At non-zero lattice spacing the direct relation between correlation functions
involving the energy operator and temperature derivatives of any observable
is violated by cut-off effects. Nonetheless we expect that these are small
in the vicinity of a second order phase transition. One thus may expect that
at least the singular behavior of correlation functions that involve 
correlations with the energy operator will be independent of Euclidean time.
Will analyze the critical behavior of $G_{\Theta\Theta} (\tau,T)$ in
the next two subsections and show explicitly that the leading singular 
behavior indeed is independent of Euclidean time.

\subsection{Finite size scaling of \boldmath$G_{\Theta\Theta}$ at the 
critical point}

We have calculated the correlation function $G_{\Theta\Theta}(\tau,T)$  
close to the deconfinement transition point of the SU(2) gauge theory. 
In most of our simulation we use lattices of size $N_\sigma^3 N_\tau$
with $N_\tau =4$. This gives us information on the correlation function
at two non-zero values of Euclidean time, {\it i.e.} at $\tau T=1/4$ and at the
midpoint $\tau T=1/2$. Most of our simulations have been performed at 
temperatures close to the phase transition where the correlation 
length becomes large. This required calculations on large
spatial lattices in order to eliminate finite volume effects. We used 
spatial lattice sizes with aspect ratios $N_\sigma/N_\tau$ varying from $8$
($32^3\times 4$ lattices) up to values as large as 32 
($128^3\times 4$ lattices). This large aspect ratio made it possible 
to perform calculations at temperatures in the proximity of $T_c$ without
being affected by large finite volume effects. We checked that
for all temperature values, except for 
calculations performed directly at $T_c$, the thermodynamic limit has been
reached within our numerical accuracy. To reach the statistical accuracy
needed for our scaling tests close to the critical point, where fluctuations
are large, we have generated a large number of gauge field configurations
ranging from $1\cdot 10^{6}$ on our smaller lattices to $4\cdot 10^{5}$
on the large lattices. 

As mentioned in Section II.B the location of the critical point is 
quite well known for lattices with temporal extent  $N_\tau = 4$
\cite{Engels:1998nv}.
The relation between gauge coupling and temperature
has been determined in the vicinity of $\beta_c$ non-perturbatively
from calculations of the transition temperature on lattices of temporal 
size $4\le N_\tau \le 16$ 
\cite{Redlich}. We use these results to determine the reduced 
temperature $t$. 

In various studies of the critical behavior of bulk thermodynamics of
the (3+1)-dimensional SU(2) gauge theory it has been established that
this gauge theory belongs to the universality class of the 3-dimensional
Ising model. In our analysis of the critical behavior of correlation 
functions of the energy-momentum tensor we use established results on the 
critical exponents $\alpha$, $\nu$ and $\omega$ as well as the universal 
ratio of the specific heat amplitudes, $A_+/A_-$. These parameters are
summarized in Table~\ref{tab:exponents}.

\begin{table}[t]
\begin{center}
\vspace{0.3cm}
\begin{tabular}{|c|c|c|c|}
\hline
~$\alpha$~&~$\nu$~&~$\omega/\nu$~&~$g_A$~ \\
\hline
~0.110(1)~ & ~0.6301(4)~ & ~0.84(4)~ & ~0.54(1)~ \\
\hline
\end{tabular}
\end{center}
\caption{Specific heat exponent ($\alpha$), correlation length
exponent($\nu$), correction-to-scaling exponent
$\omega$ and the universal ratio of amplitudes $g_A=A_+/A_-$ of the 
3-dimensional Ising model (from \cite{Pelissetto:2000ek}).}
\label{tab:exponents}
\end{table}

\begin{figure}[t]
\begin{center}
\epsfig{file=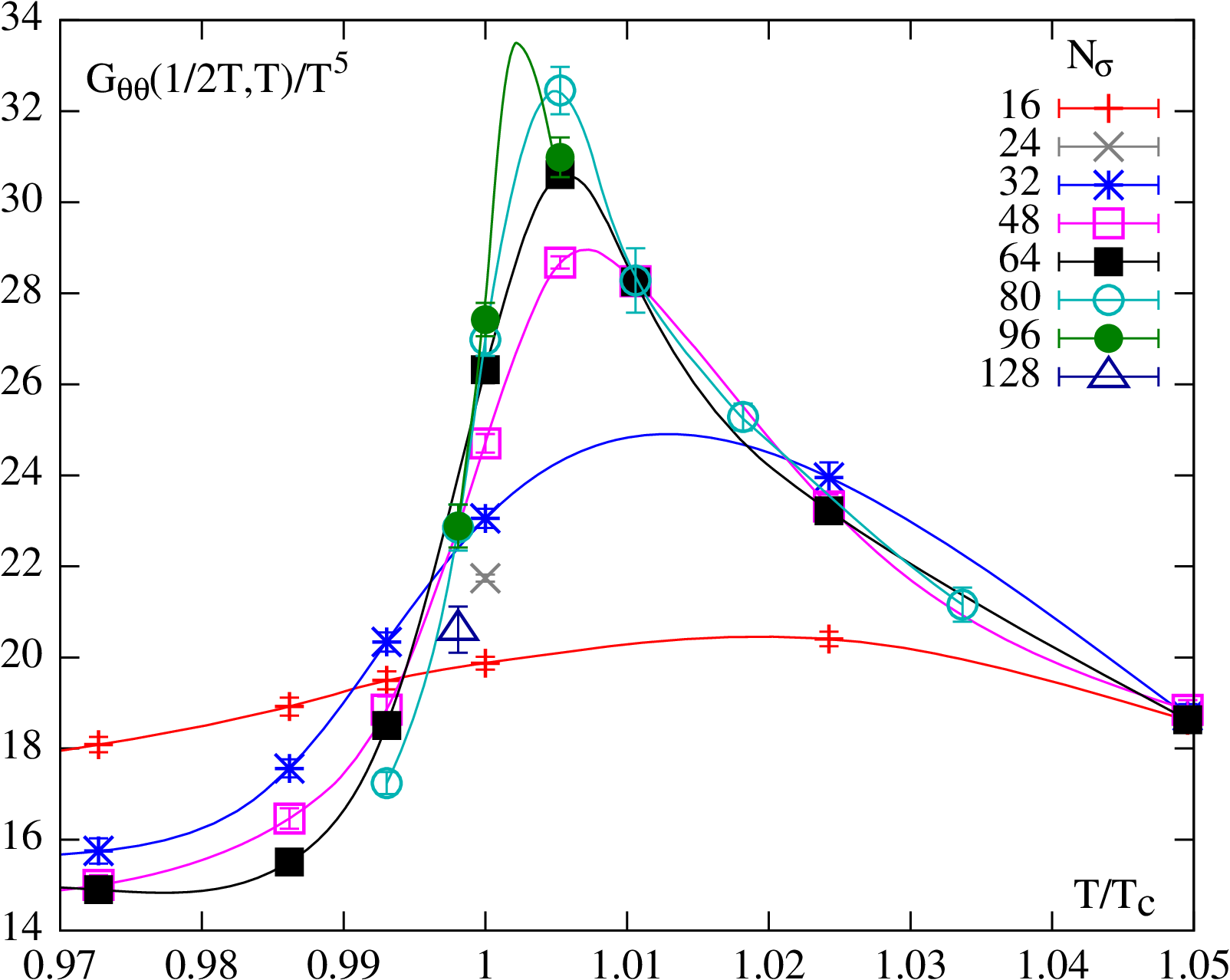, width=7.8cm}\hspace*{-0.4cm}
\epsfig{file=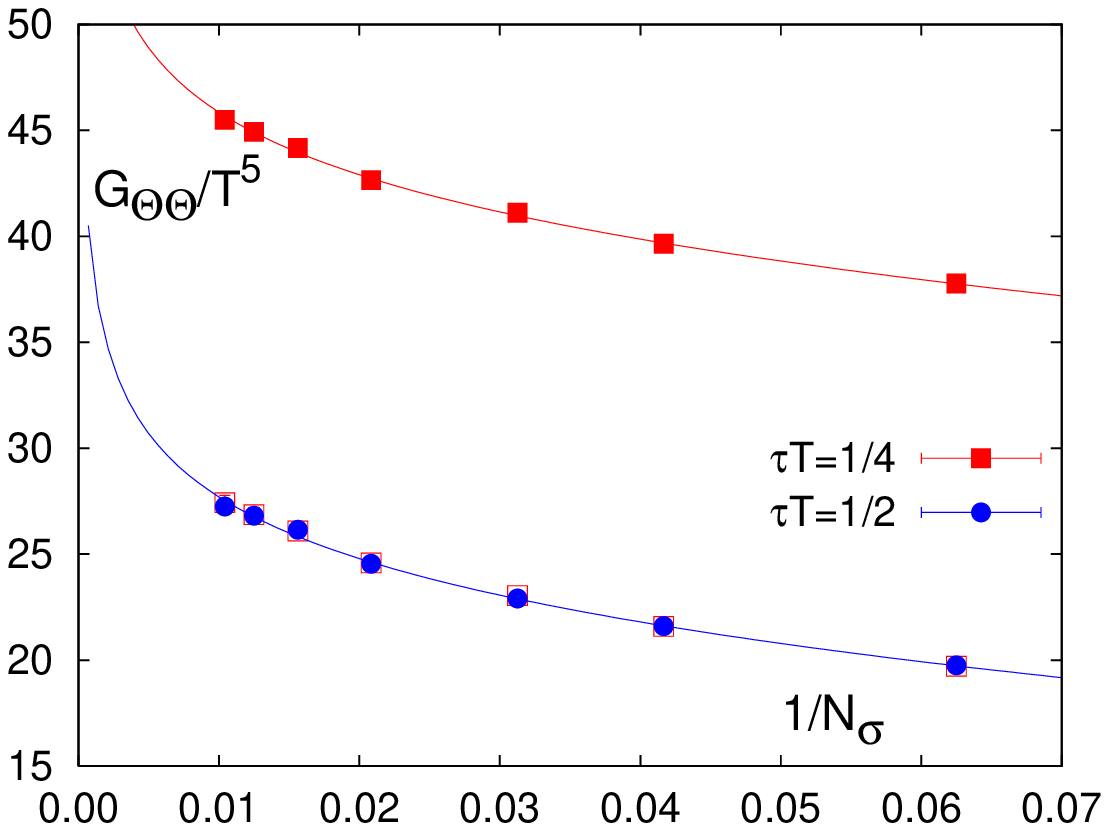, width=9.0cm}
\end{center}
\caption{The correlation function of the trace of the energy-momentum
tensor, $G_{\Theta\Theta}(\tau,T)/T^5$ at distance $\tau T= 1/2$ calculated
on lattices with temporal extent $N_\tau=4$ and  several spatial
lattice sizes $N_\sigma$ (left). The curves show spline interpolations. 
The right hand part of the figure shows the volume dependence of
$G_{\Theta\Theta}(\tau,T)/T^5$ at the critical point $T=T_c$. The curves show 
finite size scaling fits as explained in the text. The hardly 
visible open squares on top of the $\tau T=1/2$ data are the data
obtained for $\tau T=1/4$ shifted by $\Delta C_\sigma = 18.0675$.
}
\label{fig:volume}
\end{figure}

In Fig.~\ref{fig:volume}(left) we show results for the correlation function
$G_{\Theta\Theta}(\tau,T)/T^5$ calculated at the midpoint, $\tau T=1/2$,  
on lattices
with temporal extent $N_\tau =4$ and for various spatial lattice sizes. As
can be seen at all temperatures except, of course, at $T_c$ the dependence on 
the spatial volume is well under control. The peak at the pseudo-critical 
point close to $T_c$ rises only slowly with increasing size of the volume 
as the specific heat 
exponent $\alpha$ is quite small. The right hand part of Fig.~\ref{fig:volume} 
shows the volume dependence of the correlation function evaluated at $T_c$
on lattices with temporal extent $N_\tau =4$ at the two non-zero distances 
accessible in this case. At $T_c$ we have fitted 
the volume dependence of $G_{\Theta\Theta}(\tau,T_c)/T_c^5$ to a finite size
scaling ansatz, which includes the leading singular behavior of the specific heat 
and a correction-to-scaling term,
\begin{equation}
G_{\Theta\Theta}(\tau,T_c)/T_c^5 = A_\sigma N_\sigma^{\alpha/\nu} 
\left( 1 + B_\sigma N_\sigma^{-\omega/\nu} \right) + C_\sigma \; .
\label{Lscaling}
\end{equation}
Here $A_\sigma$, $B_\sigma$, $C_\sigma$ are fit parameters, which all
might depend on Euclidean time. We performed fits for the two non-zero
Euclidean time separations reachable on lattices with temporal extent $N_\tau=4$,
{\it i.e.} $\tau T= 1/4$ and $1/2$.
In the fits we make use of the known Ising exponents
$\alpha$, $\nu$ and $\omega$ which are given in Table~\ref{tab:exponents}.
This ansatz gives excellent finite size scaling fits for both data sets 
shown in 
Fig.~\ref{fig:volume}(right).  This confirms that the critical behavior is 
controlled by the exponents of the 3-dimensional Ising model. When performing
fits of the data sets at distance $\tau T = 1/4$ and $1/2$ separately, we
observe that the amplitudes $A_\sigma$ and $B_\sigma$ agree within errors.
This also is apparent from the fact that the data at $\tau T = 1/4$
can be shifted on top of the data at $\tau T = 1/2$ by a $N_\sigma$ independent
constant (open squares in Fig.~\ref{fig:volume}(right)). 
We then performed simultaneous fits of both data sets, still allowing for
different amplitudes at both distances. From this combined fit we find
that $A_\sigma(\tau T=1/2)/A_\sigma(\tau T=1/4) = 1.01\pm 0.19$ and
$B_\sigma(\tau T=1/2)/B_\sigma(\tau T=1/4) = 1.01\pm 0.58$.
The fit results\footnote{Note that the coefficients of the 
correction-to-scaling term, $B_\pm$, are
negative. Although these amplitudes are non-universal it has been noted
previously that for the 3-dimensional Ising model on various lattice
geometries this correction term comes out to be negative \cite{Fisher}.}
are summarized in Table~\ref{tab:fits}, which also includes the coefficients
extracted from an equally good combined fit of both data sets obtained requiring
the same coefficients $A_+$, $B_+$ and $B_-$ for the two Euclidean distances.

\begin{table}[t]
\begin{center}
\vspace{0.3cm}
\begin{tabular}{|c|c|c|c|c|c|c|c|c|c|}
\hline
&~$\tau T$~&~$A_+$~&~$B_+$~&~$B_-$~&~$C$~&~$D$~&~$A_\sigma$~&~$B_\sigma$~
&~$C_\sigma$~\\
\hline
free&1/4 & 16.3(2.1)& -0.59(19) & -1.9(1.4) & 21.5(5.7) & -150(53) & 9.2(1.2) & -2.4(1.0) & 26.3(2.7)\\
fit&1/2 & 16.4(2.2)& -0.76(18) & -2.2(1.5) & 3.7(3.4) & -95(32) & 9.1(1.2) & -2.4(0.9) & 8.4(2.9)\\
\hline
\hline
combined&1/4& \multirow{2}{*}{16.5(1.3)}& \multirow{2}{*}{-0.74(10)} & \multirow{2}{*}{-2.12(86)} & 21.8(2.1) & -143(20) & \multirow{2}{*}{9.15(73)} & \multirow{2}{*}{-2.39(59)} & 26.4(1.9)  \\
fit&1/2& & &  & 3.8(1.9) & -87(18) &  &  &  8.3(1.7)  \\
\hline
\end{tabular}
\end{center}
\caption{Fit parameters for fits to $G_{\Theta\Theta}(\tau,T)/T^5$ using
Eq.~\ref{ansatz} and for the volume dependence of 
$G_{\Theta\Theta}(\tau,T_c)/T_c^5$. The last row gives parameters of 
combined fits to the data for $\tau T = 1/4$ and $1/2$.
}
\label{tab:fits}
\end{table}

The analysis presented above establishes that the correlation function of the
trace of the energy-momentum tensor shows the expected universal singular 
structure of the specific heat in a 3-dimensional Ising model. 
In the vicinity of $T_c$ the singular contributions to 
$G_{\Theta\Theta}(\tau,T)$ are found to be independent of Euclidean time. 
We have performed corresponding analyzes for $N_\tau=6$ and $8$ which are 
consistent with these findings, however statistical errors are larger in
these cases. 

The analysis presented above confirms that the  correlation function
$G_{\Theta\Theta}$ contains a constant contribution that arises 
from low frequency modes that lead to the singular behavior of the
energy-energy correlation function at $T_c$. In the continuum limit
this constant will reflect the $\tau$-independence of the correlators
$G_{\epsilon\epsilon}$ and $G_{\epsilon P}$.

\subsection{Critical behavior of \boldmath$G_{\Theta\Theta} (\tau, T)$
in the vicinity of $T_c$}

In the previous subsection we established through a finite size scaling 
analysis at $T_c$ that the singular terms in 
$G_{\Theta\Theta} (\tau, T)$ are independent of Euclidean time. 
We will show here that this is also the case at temperatures in the 
vicinity of $T_c$ at which the universal scaling behavior allows to 
separate the long-distance singular behavior of $G_{\Theta\Theta} (\tau, T)$
from regular contributions.
 
\begin{figure}[t]
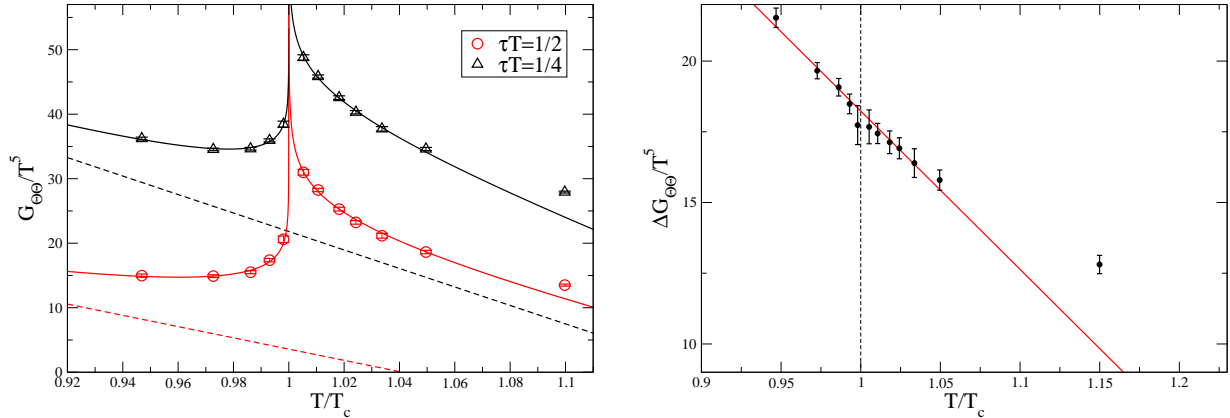

\vspace*{0.2cm}
\begin{center}
\epsfig{file=Tdep_fit.eps, width=7.8cm}\hspace*{0.5cm}
\epsfig{file=diff.eps, width=7.8cm}
\end{center}
\caption{The  correlation function of the trace of the
energy-momentum tensor, $G_{\Theta\Theta}(\tau,T)/T^5$ at distances
$\tau T= 1/4$ and $1/2$ (left).
The curve shows the combined fit of both data sets performed with the
ansatz given in Eq.~\ref{ansatz}. The dashed straight lines give the
regular part of the fit. The right hand part of this figure
shows the difference $\Delta G_{\Theta\Theta} = G_{\Theta\Theta}(1/4T,T)-
G_{\Theta\Theta}(1/2T,T)$.
}
\label{fig:scaling}
\end{figure}
 
In Fig.~\ref{fig:scaling} (left) we show infinite volume extrapolated results
for $G_{\Theta\Theta}(\tau,T)/T^5$ at two distances, $\tau T = 1/4$ and $1/2$.
Here we left out the data at $T_c$. At both distances $G_{\Theta\Theta}(\tau,T)/T^5$
increases rapidly when $T$ approaches $T_c$; above
$T_c$ the correlation function changes by a factor 2 in the temperature 
interval $[1.01\; T_c, 1.1\; T_c]$. 

In the vicinity of $T_c$ we expect 
that the data are well described by the scaling ansatz, 
\begin{equation}
G_{\Theta\Theta}(\tau,T)/T^5 = A_\pm t^{-\alpha} (1 + B_\pm t^{\omega}) + 
C + D \; t \; ,
\label{ansatz}
\end{equation}
where $A_+$, $B_\pm$, $C$, $D$ are free parameters, which again all may 
depend on Euclidean time; 
$A_+=g_A A_-$ and $\alpha$, $\omega$ and $g_A$ are the known Ising 
exponents given in Table~\ref{tab:exponents}.
We again have performed fits for the two data sets at $\tau T=1/2$ and
$1/4$ separately as well as simultaneously.
The ansatz given in Eq.~\ref{ansatz} gives good fits 
in the interval $T/T_c\in [0.94,1.05]$. As expected the parameters
$C$ extracted for $\tau T=1/2$ and $1/4$, are consistent with the constant 
terms obtained from the finite size scaling analysis at $T_c$.


Again we find that within errors the fit parameters $A_+$ and $B_\pm$ are 
independent of $\tau$; a combined fit of 
$G_{\Theta\Theta}(1/4T,T)/T^5$ and $G_{\Theta\Theta}(1/2T,T)/T^5$ with common 
amplitudes $A_+$ and $B_\pm$ also gives a $\chi^2/dof = 1.1$. The resulting fit
parameters are also given in Table~\ref{tab:fits}. 

We thus conclude that in the vicinity of  the critical point of the SU(2) 
gauge theory the temperature dependence of the correlation function
of the energy-momentum tensor reflects the singular behavior of the 
specific heat of 3-dimensional Ising model. More striking may be the fact
that the amplitudes for the singular terms are to a good approximation
independent of $\tau$. However, as discussed above this is consistent with 
the expected $\tau$-independence of the energy-energy correlation 
function \cite{Meyerc}.

In order to eliminate the leading singular behavior from 
$G_{\Theta\Theta}(\tau,T)$ it thus suffices to consider differences of the 
correlation function evaluated for different Euclidean time separations, e.g.
\begin{equation}
\frac{\Delta G_{\Theta\Theta}}{T^5} = \frac{G_{\Theta\Theta}(1/4T,T)}{T^5}-
\frac{G_{\Theta\Theta}(1/2T,T)}{T^5} \; .
\label{dG}
\end{equation}
Fig.~\ref{fig:scaling}(right) shows that this difference indeed stays finite 
at $T_c$.
However, as shown in Fig.~\ref{fig:cutoff}, at non-zero value of the 
lattice spacing $\Delta G_{\Theta\Theta}$ does not yet
coincide with $9 \Delta G_{PP}$ as it will in the continuum limit.

\begin{figure}[t]
\begin{center}
\epsfig{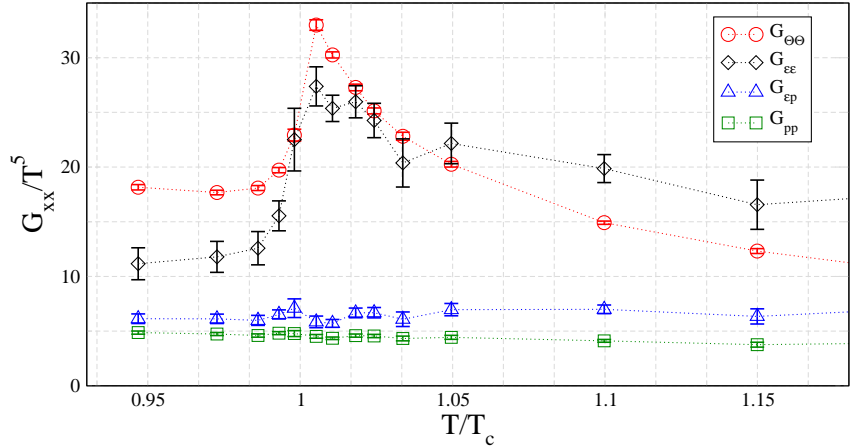}
\end{center}
\caption{Correlation functions $G_{XX}(1/2T,T)$ contributing to the correlator
of the trace anomaly. Shown are results obtained from calculations on lattices
with $N_\tau=4$ in the infinite volume limit.
}
\label{fig:Gxx}
\end{figure}

We finally show in Fig.~\ref{fig:Gxx} the temperature dependence of the 
different correlation functions, $G_{XX}(\tau, T)$, 
contributing to $G_{\Theta\Theta}$.
As can be seen the entire strong temperature dependence of $G_{\Theta\Theta}$ 
in the vicinity of $T_c$ indeed arises
from $G_{\epsilon\epsilon}$. All other correlators show only a weak 
temperature dependence in the vicinity of $T_c$. In particular, we find
that the pressure-pressure correlation function stays finite at $T_c$
and varies little in its vicinity. As will be discussed in the 
next section this has immediate consequences for the determination
of the bulk viscosity from correlation functions of the energy-momentum
tensor. 

\section{Bulk viscosity and relaxation time scale}

Correlation functions of the energy-momentum tensor are widely used
to gain information on transport coefficients. The spectral
representation of these Euclidean-time correlation functions and 
their relation to real-time Kubo formulas allow, in principle,
to extract transport coefficients from the zero frequency
limit of spectral functions \cite{Wyld}.
We want to discuss here some consequences for the calculation of 
transport coefficients, in particular the bulk viscosity, that can
be drawn from the analysis of correlation functions presented in 
the previous sections.

The bulk viscosity can be extracted from the spectral representation of
pressure-pressure correlation functions \cite{Wyld},
\begin{equation}
G_{PP}(\tau,T)= \int_0^\infty \text{d}\omega\,
\rho_{PP}(\omega,T)K(\tau,\omega,T) \; ,
\label{correlator_P}
\end{equation}
where
\begin{equation}
K(\tau,\omega,T)=
\frac{\cosh\left(\omega(\tau-\frac{1}{2T})\right)}{\sinh(\frac{\omega}{2T})}
\label{kernel}
\end{equation}
and $\rho_{PP} (\omega,T)$ is the
spectral density at vanishing momentum. Making use of the definition of
transport coefficients via the Kubo-formulas it is readily seen that the
bulk viscosity $\zeta$ is related to the low frequency limit of
this spectral function \cite{Jeon:1995zm},

\begin{equation}
\zeta(T)=\pi\lim_{\omega\to 0}\frac{\rho_{PP}(\omega,T)}{\omega}.
\label{bulk}
\end{equation}
From Eq.~\ref{decomposed} it is obvious that this can also be obtained 
from a spectral analysis of the correlator of the trace of the energy-momentum 
tensor. However, as we have verified explicitly in the previous sections this  
correlation function receives a constant contribution from correlation
functions containing the energy operator. This constant term is reflected
in zero frequency contribution to $\rho_{\Theta\Theta}$, {\it i.e.} a
$\delta$-function at vanishing frequencies,
\begin{equation}
G_{\Theta\Theta}(\tau,T)= \int_0^\infty \text{d}\omega\,
\rho_{\Theta\Theta}(\omega,T)K(\tau,\omega,T) \; ,
\label{correlator_Theta}
\end{equation}
with 
\begin{equation}
\rho_{\Theta\Theta}(\omega,T) = 9 \rho_{PP}(\omega,T) +T\partial_T(\epsilon-6P) 
\omega \delta (\omega) \; .
\label{rho_Theta}
\end{equation}
As the bulk viscosity is obtained from the spectral function in the limit of 
vanishing frequency one should eliminate the contribution arising from 
$\rho_{\Theta\Theta}(0)$.
This can, of course, easily be done by analyzing the subtracted correlation
function $\Delta G_{\Theta\Theta} (\tau,T)$ introduced in Eq.~\ref{dG}. From 
Fig.~\ref{fig:scaling}(right) we know that this correlation function will 
stay finite at $T_c$. Its spectral representation is given by
\begin{equation}
\Delta G_{\Theta\Theta}(\tau,T)= 9 \int_0^\infty \text{d}\omega\,
\rho_{PP}(\omega,T)
\frac{\cosh\left(\omega(\tau-\frac{1}{2T})\right)-1}{\sinh(\frac{\omega}{2T})}
\; .
\label{correlator_Theta_spec}
\end{equation}

The low frequency part of the spectral function, $\rho_{PP}(\omega, T)$,
is expected to be linear, the slope being proportional to the bulk 
viscosity. The rise of the spectral functions ends at a characteristic
frequency $\omega_0$, which is identified as a characteristic 
relaxation time scale, $\tau_R\equiv 1/\omega_0$, for density fluctuations. 
Of course, $\omega_0$
will be temperature dependent and, in fact, is expected to vanish at a second
order phase transition point where fluctuations grow and relaxation times 
will diverge.

We may split the spectral function $\rho_{PP}$ into a low 
frequency part and a large frequency part, $\rho_{PP}(\omega,T) \equiv
f(\omega,\omega_0) + \rho_>(\omega,T)$.
The low frequency
behavior of the spectral function may be modeled by a Breit-Wigner 
ansatz, 
\begin{eqnarray}
f(\omega,\omega_0) = \frac{9}{\pi}\zeta (T) \omega 
\frac{\omega_0^2}{\omega^2 +\omega_0^2} \quad .
\label{special_case}
\end{eqnarray} 
With this we can write the low frequency contribution to the pressure
correlator,
$G_{PP}(\tau T, T)$, as 
\begin{eqnarray}
G^{low}_{PP}(\tau, T) &=& \frac{9}{\pi}\zeta\omega_0 
\int_0^{\infty} {\rm d}x \frac{\omega_0 x}{1+x^2}
\frac{\cosh( \omega_0 x (\tau -1/2T))-1}{\sinh (\omega_0 x/2T)}  
\label{Gsing}\\
&=& 
9 T
\zeta(T)\omega_0(T) \left( 1 - \frac{\omega_0 (T)}{2T\pi} \ln\left[2 - 2 \cos (2\pi T \tau)\right] 
+ {\cal O} (\omega_0^{2}) \right) \; . \nonumber
\end{eqnarray}
Here we have assumed in the second equality that $\omega_0$ becomes
small in the vicinity of $T_c$ and an expansion in $\omega_0$ thus
is appropriate. 
Similarly we can obtain the low frequency contribution to 
$\Delta G_{\Theta\Theta} (\tau,T)$. Both correlation functions have the same 
functional dependence on Euclidean time for $\tau T\simeq 1/2$. It is 
this region, where one can hope that the low frequency part of the
spectral function dominates the $\tau$-dependence of the 
correlation functions and thus will allow a determination of transport
properties. 

From Eq.~\ref{Gsing} we find that the low frequency part of the spectral
function, which is proportional to bulk viscosity, contributes a term
proportional to $\zeta \omega_0$ to the value of the correlation function
at the midpoint, $\tau T=1/2$. As discussed in the introduction,
universality arguments given for the scaling of the bulk viscosity
in the vicinity of a critical point in the Ising universality class
suggest, however, that the ratio of bulk viscosity and relaxation time,
$\zeta/\tau_R \equiv \zeta \omega_0$, vanishes as the inverse of the 
specific heat, {\it i.e.} it is expected to be proportional to the 
velocity of sound. In the vicinity of $T_c$ the correlation function
$G_{\Theta\Theta}(1/2T, T)$ thus is dominated by contributions from
the large frequency, regular part of the spectral function. The 
sensitivity to bulk viscosity also is small away from the midpoint 
where contributions from the low frequency part of the spectral
function are proportional to $\zeta \omega_0^2$. To separate this
contribution from the large frequency contributions it would be necessary
to become sensitive to the characteristic logarithmic dependence on
Euclidean time that arises from the low frequency part of the spectral
function.   

\section{Conclusions}

We have shown that the correlation function of the trace of the energy-momentum
tensor, $G_{\Theta\Theta}$, diverges at the critical point of the finite 
temperature SU(2) gauge
theory in (3+1)-dimensions. The divergence is controlled by the critical
exponent $\alpha$ of the 3-dimensional Ising model that also controls the 
divergence
of the specific heat. Furthermore, we have shown that $G_{\Theta\Theta}$ becomes
independent of Euclidean time at the critical point, which indicates that
its spectral representation receives a contribution from a  $\delta$-function 
at zero frequency. This confirms the observations made in Ref.~\cite{Meyerc}.

To get direct access to the bulk viscosity $\zeta$
and the characteristic frequency $\omega_0$ that controls the relaxation
time for density fluctuations will require much more detailed studies of the
dependence of correlation functions on Euclidean time. In fact, we conclude
that a determination of bulk viscosity from correlation functions of the 
energy-momentum tensor is not possible without a simultaneous determination
of the relevant relaxation time scale.  

\section*{Acknowledgments}
\label{ackn}
We thank Harvey Meyer, Dimitri Kharzeev and Kirill Tuchin for 
helpful discussions.
This work has been supported by the contract DE-AC02-98CH10886
with the U.S. Department of Energy.
Numerical simulations have been performed on the BlueGene/L
at the New York Center for Computational Science (NYCCS).

\end{document}